\documentclass[12pt]{article}
\begin{document}

\begin{center}
{\bf Calculation and modular properties of  multi-loop
superstring amplitudes} \\
G. S. Danilov\\ Petersburg Nuclear Physics Institute,
Gatchina, Russia\\
E-mail: danilov@thd.pnpi.spb.ru
\end{center}
\begin{center}
{\bf Abstract}
\end{center}

Multi-loop
superstring amplitude are calculated in the conventional gauge where
Grassmann moduli are carried by the 2D gravitino field.
Generally, instead of the modular symmetry, the amplitudes hold the
symmetry under modular transformations added by relevant
transformations of the 2D local supersymmetry.  If a number of loops is
larger than $3$, the integration measures are not modular forms. In
this case the expression for the amplitude contains an integral over
the bound of the fundamental region of  the modular group.

\section{Introduction}

In the Ramond-Neve{\'u}-Schwarz
theory the world sheet is often specified \cite{ver}
as the Riemann surface with spin structure
\cite{swit}.  The spin
structures are not invariant under
transformations of the 2D supersymmetry. It
leads to the well known difficulties
\cite{ver,momor,as} in the calculation of
the multi-loop interaction amplitudes.
They depend \cite{ver}
on the 2D gravitino field \cite{momor,as}. It means that the
world-sheet supersymmetry is lost. Indeed, in the superstring theory the
vierbein and the 2D-gravitino field  are
the  gauge fields of the group of local symmetries on the
string world sheet.  Owing to gauge invariance, the
``true'' amplitudes are independent of the choice of the gauge of the
above fields.

In \cite{hok} true two-loop amplitudes have been
obtained.  The calculation of the Ramond-Neve{\'u}-Schwarz amplitudes
with any number of loops  has been done \cite{danphr,dan04} in the
supercovariant  scheme \cite{bsh,vec,dan93}.  In this case the zweibein
and 2D gravitino field are conformally flat.  The string
world sheet is specified as the $(1|1)$ complex, non-split
supermanifold. The supermanifold  carries a ``superspin'' structure
\cite{danphr,dan93,dannph} instead of the spin structure \cite{swit}.
The superstring amplitude is obtained by a
summation over the superspin structures.
The superspin structures are supersymmetric extensions
of the spin structures \cite{swit}.
In this case the twist about $(A,B)$-cycles
is, generally, accompanied by a supersymmetric
transformation including fermion-boson mixing.
The fermion-boson mixing
arises due to the presence of Grassmann moduli that are assigned to the
$(1|1)$ complex, non-split
supermanifold in addition to the Riemann ones.
The fermion-boson mixing differentiates the
superspin structures from the ordinary spin ones. Indeed, the ordinary
spin structures \cite{swit} imply that boson fields are single-valued
on Riemann surfaces. Only fermion fields being twisted about
$(A,B)$-cycles, may receive the sign.  The $g$-loop
spinning (fermion) string interaction amplitude (with $g>1$) is given
by an integral over $(3g-3|2g-2)$ complex moduli and over interaction
vertex coordinates on the supermanifold. The integrand (the local
amplitude) has been explicitly  calculated \cite{danphr} for every
superspin structure. The calculation employes only the gauge symmetry
of the fermion string. In doing so, the partition functions are
computed from equations
\cite{danphr,dan90} that are nothing else than  Ward identities. These
equations  realize the requirement that the superstring amplitudes
are independent of both the vierbein and the gravitino field.
Therefore, the obtained multiloop amplitudes
are consistent with the gauge invariance of the superstring theory.
The world-sheet gauge group is so large that  the Ward identities
fix the partition function up to a constant factor.
This factor is determined by the factorization condition following from
the unitarity equations. The module space integral of the local
amplitude is, however, ambiguous \cite{as,dan04} under non-split
replacements of the moduli.  The ambiguity is also present in the sum
over super-structures for the superstring amplitude.  The
ambiguity in the superstring amplitude is resolved \cite{dan04} so that
the cosmological
constant is equal to zero. The obtained superstring amplitudes are
finite.  The one-, two- and three-point massless boson
amplitudes vanish in accordance with the space-time supersymmetry.

In this paper, we consider the  calculation of the same multi-loop
amplitudes in the above-mentioned gauge \cite{ver} where the $(1|1)$
complex supermanifold is split, in the sense that fermions are not
mixed to bosons under twists around non-contractible cycles.  The
genus-$g$ supermanifold is specified as the genus-$g$ Riemann surface
$\Sigma_g$ with the given spin structure.  The spinning string
amplitudes each is represented by an integral of a local amplitude
where the integration is performed over vertex coordinates and over
the moduli.  The superstring amplitude is obtained  by  summing over
spin structures.  Grassmann moduli are carried by the 2D-gravitino
field $\phi_m$ which is usually specified such that $\gamma^m\phi_m=0$
where $\gamma^m$ is the Dirac 2D-matrix.  Non-zero components
$\phi_{\pm}(z,\bar z)$ of the $\phi_m$ field  are given by
\begin{equation}
\phi_-(z,\bar
z)=\sum_{s=1}^{2g-2}\lambda_s\phi_{s-}(z,\bar z)\,, \quad
\phi_+(z,\bar z)=
\sum_{s=1}^{2g-2}\overline\lambda_s\phi_{s+}(z,\bar z)\,,
\label{grfield}
\end{equation}
$\lambda_s$ and $\overline\lambda_s$ being Grassmann moduli, and
$\phi_{s\mp}(z,\bar z)$ fields may depend on the Riemann moduli too.
As far as the world-sheet supermanifold is split,
one seemingly  avoids the ambiguities \cite{as,dan04}
complicating the calculation of the amplitudes in the superconformal
gauge.  Also, it seems that the amplitudes might possess the modular
symmetry and be represented through theta-functions and modular forms.
Indeed, in the two-loop calculation \cite{hok} the genus-2 integration
measures are modular forms, and the GSO projections of the
local amplitudes with less than four legs are  equal to zero. The
GSO projection of the four-point local amplitude does not depend on
$\phi_{s\mp}$. The spinning string amplitude ceases to depend on
$\phi_{s\mp}$ due to the integration over vertex coordinates.  The
papers \cite{hok} have initiated the efforts \cite{capige} to build
genus $g>2$ amplitudes assuming certain properties of the amplitudes,
the modular symmetry being among them.  This strategy meets with
difficulties \cite{dunmorsl,ma2vo}, at least for $g>3$.

The calculation of the multi-loop interaction  amplitudes in the
present paper is similar to the calculation
\cite{danphr}  in the supersymmetric gauge.  It exploits the gauge
symmetry on the string world sheet and uses no assumptions.
In this case
the spinning string amplitudes are
independent of
local variations of the $\phi_{s\mp}$ fields, but the
local amplitudes depend on $\phi_{s\mp}$ (the last takes place even in
the two-loop case \cite{hok}). Integration of the local amplitude over
the moduli and over vertex coordinates is performed at fixed
$\phi_{s\mp}$.  The integral must be invariant under re-definitions of
the non-contractable cycles on the string world sheet. The
re-definitions of the non-contractable cycles are accomplished  by
modular transformations, but these transformations, generally, change
$\phi_{s\mp}$.  Returning back to the original $\phi_{s\mp}$ is
achieved by an extra transformation of a local 2D supersymmetry.
Therefore, the symmetry group of the amplitude consists of modular
transformations accompanied by the relevant supersymmetry ones.  These
supermodular transformations are conveniently discussed in the
supersymmetric description \cite{dan90} of the fermion string on the
$(1|1)$ complex supermanifold. The period matrix
\cite{danphr,dan04,dannph} on the above-mentioned supermanifold
collects periods of scalar superfunctions which vanish under the
supercovariant Laplacian \cite{dan04}.  The supermodular transformation
changes \cite{danphr,dan04,dannph} this matrix
just as the relevant modular transformation changes the period matrix
\cite{siegal} on the Riemann surface $\Sigma_g$.

A loss of the supersymmetry in \cite{ver} occurs because
the difference between the supermodular  and  modular symmetries was
ignored and, also, because of  an incomplete calculation of the
ghost zero mode contribution to the integration measure, see
Sec.  2 and Sec. 3 below.

Since the superscalar functions depend on  $\phi_m$,
the period matrix on the $(1|1)$ complex supermanifold is, generally,
distinguished from the period matrix  on $\Sigma_g$ by  terms
proportional to the Grassmann moduli. In this case the integration of
the local amplitude over the fundamental region of the modular group
leads to the loss of 2D supersymmetry.  As the result, the spinning
string amplitudes depend on $\phi_{s\mp}$. To restore
the  supersymmetry, the integration over the fundamental
region of the modular group
must be supplemented \cite{dan04} by the integral around the
boundary of the region.  If $g\leq3$,
the periods of the superscalar functions can be taken \cite{hok}  as
the moduli set.  In this case the boundary integral does not arise.  If
$g\leq3$ and the moduli setting \cite{ver} is used, the boundary
integral is removed (see Sec. 4) by a re-definition of the
local amplitude.  The integration measures are given by modular forms
for both $g=2$ and $g=3$.  Unlike the two-loop case, the GSO projection
of the four-point, three-loop amplitudes ceases to depend on
$\phi_{\mp}$ due to the integration over vertex coordinates, just as it
arises in each of the spin structures.  If $g>3$, periods of
superscalar functions depend on Grassmann moduli for any choice of
moduli variables. The boundary integral is  present in the expression
for the amplitude, and the integration measures are not modular forms.
It is akin to what occurs in the superconformal gauge.  Hence the
strategy \cite{capige} is not in accord with the 2D supersymmetry.

In Sec. 2 the integration over moduli is discussed. In Sec. 3 local
amplitudes are calculated. In Sec. 4 two- and three-loop
amplitudes are considered in more details.

\section{Integration of local amplitudes}

As  noted in the Introduction, the period matrix on
the $(1|1)$ complex supermanifold determines
the periods
of the scalar superfunctions. The scalar superfunctions vanish
under
the super-Laplacian
$(D_+^{(\phi)T}D_-^{(\phi)}-D_-^{(\phi)T}D_+^{(\phi)})/2$ where
$T$ denotes transposing.
Operators
$D_+^{(\phi)}$ and $D_-^{(\phi)}$ depend on the gravitino
field (\ref{grfield}). We assume that the $\phi_{s\mp}$
fields do not
overlap.  Then \cite{dan04}
\begin{eqnarray}
D_-^{(\phi)}=D+\frac{1}{2}\phi_+(z,\bar
z) \biggl[\vartheta\frac{\partial}{ \partial\overline\vartheta}+
\overline\vartheta\vartheta\frac{\partial}{
\partial\overline z}\biggl]\,,\quad
D=\vartheta\partial_z+\partial_\vartheta\,,
\nonumber\\
D_+^{(\phi)}=\overline{D}
+\frac{1}{2}\phi_-(z,\bar z)
\biggl[\overline\vartheta\frac{\partial}{
\partial\vartheta}-
\overline\vartheta\vartheta\frac{\partial}{
\partial z}\biggl]
\label{difsop}
\end{eqnarray}
where
$\vartheta$ is the superpartner of $z$.
The scalar superfunctions ${\cal J}_r^{(R\sigma)}(z,\bar
z,\vartheta)$ are associated with the right movers,
and ${\cal J}_r^{(L\sigma')}(z,\bar z,\bar\vartheta)$
are associated with the left movers;
$\sigma$ and
$\sigma'$  mark spin structures.
The desired superfunctions are represented as
\begin{equation}
{\cal J}_r^{(R\sigma)}(z,\bar z,\vartheta)=J_r^{(R\sigma)}(z,\bar z)+
\vartheta\eta_r^{(R\sigma)}(z,\bar z)\,,\quad
{\cal J}_r^{(L\sigma')}(z,\bar z,\bar\vartheta)=
J_r^{(L\sigma')}(\bar z,z)+
\overline\vartheta\eta_r^{(L\sigma')}(\bar z,z)\,.
\label{ssfuns}
\end{equation}
They
can be found from the equations
\begin{equation}
D_+^{(\phi)}{\cal J}_r^{(R\sigma)}(z,\bar z,\vartheta)=0\,,\quad
D_-^{(\phi)}{\cal J}_r^{(L\sigma')}(z,\bar z,\bar\vartheta)=0\,.
\label{zereq}
\end{equation}
If $\phi_+(z,\bar z)=\phi_-(z,\bar z)\equiv0$, then
$J_r^{(R\sigma)}(z)$ is reduced to the scalar function
$J_r(z,\bar z)$, and $J_r^{(L\sigma)}(z,\bar z)$ is reduced
to\footnote{Throughout the paper, the line over denotes
complex conjugate} $\overline{J_r(z)}$.
Under $2\pi$-twists about  $B$-cycles on the
Riemann surface $\Sigma_g$
(which specifies the genus-$g$ supermanifold in
question) the $J_r^{(R\sigma)}(z,\bar z)$ functions receive periods
forming the $\Omega^{(R\sigma)}$ matrix.
Correspondingly, the periods of
$J_r^{(L\sigma')}(z,\bar z)$ form the $\Omega^{(L\sigma')}$
matrix. In this case
\begin{equation}
\Omega^{(R\sigma)}=
\Omega+\widetilde\Omega^{(R\sigma)}\,,
\quad\Omega^{(L\sigma')}=
\overline\Omega+\widetilde\Omega^{(L\sigma')}
\label{matper}
\end{equation}
where $\widetilde\Omega^{(R\sigma)}$ and
$\widetilde\Omega^{(L\sigma')}$ vanish when all the Grassmann
moduli are equal to zero, and $\Omega$ is the period matrix on
$\Sigma_g$. Eqs. (\ref{zereq})  can be transformed to
the integral equations.  In doing so, the
desired equations for the
$(J_r^{(R\sigma)}(z,\bar z),\eta_r^{(R\sigma)}(z,\bar z))$ pair and for
$\Omega^{(R\sigma)}_{sr}$ elements  of $\Omega^{(R\sigma)}$-matrix
are found to be
\begin{eqnarray}
J_r^{(R\sigma)}(z,\bar
z)=J_r(z)-\frac{1}{2\pi}\int \partial_{z'}\ln[E(z,z')]\,\phi_-(z',\bar
z') \eta_r^{(R\sigma)}(z',\bar z')d^2z'+const\,,
\nonumber\\
\eta_r^{(R\sigma)}(z,\bar z)=\frac{1}{2\pi}\int
S_\sigma(z,z')\phi_-(z',\bar z')
\partial_{z'}J_r^{(R\sigma)}(z',\bar z')d^2z\,,
\nonumber\\
\Omega^{(R\sigma)}_{sr}=\Omega_{sr}-i\int
\partial_{z}J_r(z)\,\phi_-(z,\bar z)
\eta_r^{(R\sigma)}(z,\bar z)d^2z
\label{inteq}
\end{eqnarray}
where
$J_r(z)$ is the scalar function on $\Sigma_g$ having the $\Omega_{nr}$
periods, $E(z,z')$ is the prime form and $S_\sigma(z,z')$ is the Szego
kernel \cite{ver,fay}. For the even spin structure
${\bf\sigma}=({\bf\sigma_1},{\bf\sigma_2})$ it is given by\cite{ver}
\begin{eqnarray}
S_\sigma(z,z')=\frac{\theta[\sigma]({\bf
z}-{\bf z}')}{E(z,z') \theta[\sigma](0)}\,,\quad {\bf z}-{\bf
z}'=\int_{z}^{z'}{\bf v}(x)dx\,,\quad{\bf v}=\{v_s(x)\},\quad
v_s(x)=\partial_xJ_s(x)\,,
\nonumber\\
\theta[\sigma]({\bf
z})=\theta({\bf z}+\Omega{\bf\sigma_1}+{\bf\sigma_2})
\exp[i\pi{\bf\sigma_1}\Omega{\bf\sigma_1}+2\pi i
{\bf\sigma_1}({\bf z}+{\bf\sigma_2})]\,,\quad
{\bf\sigma_i}=\{\sigma_{is}\},\quad 1\leq s\leq g
\label{ffg}
\end{eqnarray}
where ${\bf z}$ is related to $z\subset\Sigma_g$  by
the Jacobi mapping.  Further on, $\theta[\sigma]({\bf z})$  is
\cite{ver,fay} the theta-function with characteristics
${\bf\sigma}=({\bf\sigma_1},{\bf\sigma_2})$ corresponding to
the spin structure $\sigma$. The  $\theta[\sigma]({\bf z})$ function is
related with the Riemann theta function $\theta({\bf z})$ as it is
seen from the second line of (\ref{ffg}).
The first two equations in (\ref{inteq})
are equivalent  to the first
equation in (\ref{zereq}).   To verify it,
every integral equation reduced to the differential equation
by the
$\overline\partial_z$ operator.
Indeed,
$S_\sigma(z,z')\to1/(z-z')$,
$\partial_{z'}\ln[E(z,z')]\to-1/(z-z')$ at $z\to z'$, and
$\partial_{\bar z}1/z=\pi\delta^2(z)$ where
$\delta^2(z)\equiv\delta(Re\,z)\delta(Im\,z)$.
Therefore,
the integration equations (\ref{zereq}) are reduced to the differential
ones.  Due to (\ref{difsop}), these differential
equations are identical to the first equation  in
(\ref{zereq}).  The third equation in (\ref{inteq}) determines
the periods of $J_r^{(R\sigma)}(z,\bar
z)$. It is obtained  from the first equation by means of the
transformation which is assigned to the $2\pi$-twist about the
$B_s$-cycle on $\Sigma_g$.  Since the kernels in (\ref{inteq}) are
proportional to the Grassmann moduli, eqs.(\ref{inteq}) are solved by
the iteration procedure.  The functions and the period
matrix for the left movers  are calculated in the similar
manner.

Thus
$\Omega^{(R\sigma)}$ and $\Omega^{(L\sigma')}$ depend on Grassmann
moduli. Under these conditions, the moduli space integral over the
fundamental region of the modular group \cite{siegal} is not invariant
under the supermodular transformations that leads loss of the 2D
supersymmetry.
To restore the supermodular symmetry, the discussed integral
is supplemented by an
integral over the boundary of the integration region. To derive this
boundary integral, it is useful to define a function which is an
extension of the step function $\rho(x)$
(being $\rho(x)=1$ for $x>0$, and $\rho(x)=0$ for $x<0$) to the case
when $x=x_b+x_s$ contains the ``soul'' part $x_s$ that is the
part proportional to the Grassmann parameters.  Then  $\rho(x)$ is
understood in the sense that it is the Taylor series in $x_s$. In the
calculation of the Taylor series one employs the known relation
$d\,\rho(x_b)/d\,x_b=\delta(x_b)$ where $\delta(x)$ is the Dirac
delta-function, and the ``body'' $x_b$ of $x$ contains no Grassmann
parameters.  Under this convention  the fermion string interaction
amplitude $A_{\sigma,\sigma'}$  can be represented as the integral of
the ${\cal A}_{\sigma,\sigma'}$  local amplitude as follows
\begin{equation}
A_{\sigma,\sigma'}=\int{\cal
A}_{\sigma,\sigma'}\,{\cal O}(\Omega^{(R\sigma)},\Omega^{(L\sigma')})\,
\biggl(\prod_i\widetilde{\cal O}(z_i,\bar z_i;q,\bar q)\, d^2z_i
\biggl)\, d^2q\,d^2\lambda
\label{main}
\end{equation}
where
${\cal
A}_{\sigma,\sigma'}$  depends on the Riemann
moduli $(q,\bar q)$, the Grassmann moduli
$(\lambda,\bar\lambda)$ and on the $(z_i,\bar
z_i)$ coordinates of the $i$-th interaction vertex.
In this case $q=\{q_n\}$ and
$\lambda=\{\lambda_j\}$. Generally, $\Omega^{(R\sigma)}
\equiv\Omega^{(R\sigma)}(q,\lambda)$ and $\Omega^{(L\sigma')}
\equiv\Omega^{(L\sigma')}(\bar q,\bar \lambda)$. The
$\widetilde{\cal O}_j(z_j,\bar z_j;q,\bar q)$ factor is a step
function product restricting
the integration region (that is the fundamental region of the Klein
group) on the complex $z_i$-planes.  The integration region over the
moduli space is ``restricted'' by the ${\cal
O}(\Omega^{(R\sigma)},\Omega^{(L\sigma')})$ step function product.
In this case
\begin{equation}
{\cal
O}(\Omega^{(R\sigma)},\Omega^{(L\sigma')})=
\prod_j\rho({\cal G}_j)\,, \quad{\cal
G}_j\equiv {\cal G}_j(\Omega^{(R\sigma)},\Omega^{(L\sigma')})\,.
\label{bound}
\end{equation}
The set of the
${\cal
G}_i(\Omega,\overline\Omega)=0$ conditions gives the boundary
\cite{siegal} of
the fundamental region of the modular group.
The step functions $\rho({\cal G}_j)$ in (\ref{bound}) are treated as
the Taylor series in $\widetilde\Omega^{(R\sigma)}_{pq}$ and in
$\widetilde\Omega^{(L\sigma')}_{pq}$ matrix elements. Therefore,
\begin{equation}
{\cal O}(\Omega^{(R\sigma)},\Omega^{(L\sigma')})
={\cal B}_{R\sigma}{\cal B}_{L\sigma'}
{\cal O}(\Omega,\overline\Omega)
\label{sbound}
\end{equation}
where the differential operators
${\cal B}_{R\sigma}$ and ${\cal B}_{L\sigma'}$
are  defined as follows
\begin{eqnarray}
{\cal B}_{R\sigma}=1+
\sum_{p\leq q}\widetilde\Omega^{(R\sigma)}_{pq}
\frac{\partial}{\partial\Omega_{pq}}+
\frac{1}{2}
\sum_{p\leq q}\sum_{r\leq s}
\widetilde\Omega^{(R\sigma)}_{pq}\widetilde\Omega^{(R\sigma)}_{rs}
\frac{\partial}{\partial\Omega_{pq}}
\frac{\partial}{\partial\Omega_{rs}}+\dots\,,
\nonumber\\
{\cal B}_{L\sigma'}=
1+\sum_{p\leq q}
\widetilde\Omega^{(L\sigma')}_{pq}
\frac{\partial}{\partial\overline\Omega_{pq}}+
\frac{1}{2}\sum_{p\leq q}\sum_{r\leq s}
\widetilde\Omega^{(L\sigma')}_{pq}\widetilde\Omega^{(L\sigma')}_{rs}
\frac{\partial}{\partial\overline\Omega_{pq}}
\frac{\partial}{\partial\overline\Omega_{rs}}+\dots\,.
\label{omod}
\end{eqnarray}
The derivatives in (\ref{omod}) are calculated assuming that the period
matrix elements are unrelated to each other up to the transposing
operation.  Under the operators (\ref{omod}), the step function in
(\ref{sbound}) receives $\delta$-function-type terms that leads to the
appearances of the integral over the
boundary of the fundamental region of the modular group.

Under  the change of integration variables in
(\ref{main}), the arguments of
the step functions are  correspondingly replaced. As the result the
amplitude (\ref{main}) is independent of the choice of the
integration variables.  Strictly speaking, the last statement implies
that the integral (\ref{main}) is properly regularized at the points
where the Riemann surface is degenerate, but we do not discuss this
matter in the present paper. The amplitude (\ref{main}) can be derived
\cite{dan04} by a change of integration variables in the
expression for the same amplitude in the superconformal gauge
\cite{bsh,vec}. Hence (\ref{main}) is independent of $\phi_{s\mp}$
(we have directly verified it for $g\leq3$).

\section{Local amplitude ${\cal
A}_{\sigma,\sigma'}$}

To derive ${\cal
A}_{\sigma,\sigma'}$ in (\ref{main}),  we start
\cite{danphr,dan04,dan90} with the
integral \cite{pol} over all the fields, including
the zweibein  and the world-sheet
gravitino field.
The integral is divided by the volume of the local
group ${\cal
G}$ of the world-sheet symmetries of the  fermion string.
So far as  the
zweibein and the world-sheet gravitino field are
arbitrary, we can map \cite{dan90,danphr} the Riemann
surface onto the complex plane $w$ choosing the same transition group
$\widehat{\cal G}_t$
for all surfaces of the given
genus-$n$. There is no integration over any
moduli.
The zweibein and gravitino
fields can be reduced to the full set of the reference fields.
It is performed by
globally defined transformations of the ${\cal
G}$ group that do not change the $\widehat{\cal G}_t$
transition group.
The reduction is impossible within the
full set of the reference fields.
The  zweibein and the gravitino
field are represented in terms of
the  reference fields and of the gauge functions. Since the gauge
functions correspond to the ${\cal
G}$ group transformations, the $\widehat{\cal G}_t$
transition group is unchanged  and, therefore, it is the same
for all the genus-$g$ surfaces. In this case the reference fields
(for $g>1$) depend on $(3g-3|2g-2)$ complex moduli (defined up to the
supermodular transformations). Locally,  the reference fields are
arbitrary.  The integration over the zweibein and the gravitino field
is transformed to the integration over the gauge functions and
the moduli.
In doing so the Jacobian of the transformation is represented by the
integral over the ghost fields and over $(3g-3|2g-2)$ global  complex
variables dual to the
$(3g-3|2g-2)$ complex moduli. Calculating alterations of the integral
under infinitesimal local variations of the reference fields, one can
derive the Ward identities \cite{danphr,dan04} from the condition that
the amplitude (\ref{main}) is unchanged under the above-mentioned
variations of the reference fields. The obtained Ward identities are
transformed to the desirable gauge of the reference fields.
The Ward identities can be used for the calculation of the
local amplitude. Indeed, the direct calculation the amplitude from the
integral over the fields is hampered as determinants of the
differential operators appear in the calculation.  Therefore, the
integral requires a regularization ensuring the independence of the
amplitude (\ref{main}) from infinitesimal local variations of the
reference fields.  In the considered gauge \cite{ver} the discussed
uncertainty is, however, appears in the local
amplitude as the
factor which is independent of the Grassmann moduli and of
the vertex coordinates.  Excepting this factor, the local amplitude can
be obtained from the discussed integral over the fields and over the
$(3g-3|2g-2)$ global complex variables $(\Lambda_m^{b}|\Lambda_s^{f})$
dual to the $(3g-3|2g-2)$ complex moduli. The integral is, as follows
\cite{danphr,dan04,dan90}
\begin{equation}
{\cal
A}_{\sigma,\sigma'}={\cal A}_{\sigma,\sigma'}^{(v)}<\prod_jV_j>_\phi=
\int(D{\cal F})d^2\Lambda^b
d^2\Lambda^f\,(V)\exp[S_m+ S_{gh}^{R}+
S_\Lambda^{R}+S_{gh}^{L}+S_\Lambda^{L}]
\label{ammain}
\end{equation}
where ${\cal A}_{\sigma,\sigma'}^{(v)}$ is a
``vacuum local amplitude'' and
$<\prod_jV_j>_\phi$ is the vacuum expectation
of
the vertex product $(V)$. The vacuum expectation is calculated in
the gravitino field (\ref{grfield}).
Further, $(D{\cal F})$ is the product of
differentials of the fields, the fields  being
the ghost complex fields  and
10 scalar $x^N$ fields with their superpartners
$(\psi^N,\overline\psi^N)$.
The ghost complex fields are $(2,-1)$-tensor fields $(b,c)$ and
$(3/2,-1/2)$-tensor fields $(\beta,\gamma)$.  Further on,
\begin{eqnarray}
\widetilde
S_m=\frac{2}{\pi}\sum_N\int\,d^2z\biggl[-\overline\partial x^N \partial
x_N\,+\psi^N\overline\partial\psi_N\,+\overline\psi^N
\partial\overline\psi_N+\phi_-\psi^N\partial x_N+
\phi_+\overline\psi^N\overline\partial x_N\biggl]\,,
\label{mact}\\
S_{gh}^{R}=\frac{1}{\pi}\int\,d^2z\biggl[
-b\overline\partial c+
\beta\overline\partial\gamma-\frac{1}{2}\phi_-(b\gamma+\beta
\partial c)+\beta (\partial\phi_-)c\biggl]\,,
\label{ghact}\\
S_\Lambda^{R}=
\frac{1}{\pi}\int\,d^2z\biggl[-\sum_m\biggl(b\overline\partial
\varsigma_m +\frac{1}{2}\phi_-\beta \partial \varsigma_m-\beta
\partial\phi_-\varsigma_m-
\beta\frac{\partial\phi_-}{\partial q_m}\biggl)\Lambda_m^{b}
+
\sum_s\beta\phi_{s-}\Lambda_s^{f}\biggl]
\label{lact}
\end{eqnarray}
where $\partial
f\equiv\partial_zf$ and $\bar\partial
f\equiv\overline{\partial_z}f$
for any function $f\equiv f(z,\bar z)$. Also,
$\phi_{s-}\equiv\phi_{s-}(z,\bar z)$ and
$\phi_-\equiv\phi_-(z,\bar z)$, see
(\ref{grfield}).
The $S_{gh}^{L}$ and  $S_\Lambda^{L}$
in (\ref{ammain})
are obtained by the complex conjugation of (\ref{ghact}) and
of (\ref{lact}) together with the
$\overline{\phi_-(z,\bar z)}\to\phi_+(z,\bar z)$
replacement.

Eq.(\ref{lact}) contains a function
$\varsigma_m\equiv \varsigma_m(z,\bar z)$ that  is
one-valued
under rounds about $A$-cycles and
has a discontinuity under twists
about $B$-cycles on the Riemann surface. Let assign the $z\to g(z)$
replacement\footnote{For the sake of simplicity we assume the Schottky
description of the Klein group, but the using of the Schottky moduli
is not implied.}
to the $2\pi$ twist about $B_s$-cycle. Then
\begin{equation}
\varsigma_m(g_s(z),\overline{g_s(z)})=
\varsigma_m(z,\overline z)
\frac{\partial g_s(z)}{\partial z}
+\frac{\partial g_s(z)}{\partial q_m}\,.
\label{jump}
\end{equation}
The last term on the right side
of (\ref{jump}) is the discontinuity. Due to the
discontinuity in $\varsigma_m(z,\overline z)$, the integration  over
zero modes of $b$-fields in (\ref{ammain}) is convergent.
As explained below, the result of the integration
in (\ref{ammain}) does not depend on a further specification of
$\varsigma_m(z,\overline z)$.

The $V$ vertex in (\ref{ammain}) is built using the
supercovariant operators (\ref{difsop}), but it can be verified that
the $\phi_{\mp}$-dependent terms in (\ref{difsop}) do not contribute to
the amplitude. It appears due to motion equations following from
(\ref{mact}).  So the conventional vertex \cite{fried} can be employed.

With the exception of the $(\lambda_s,\bar\lambda_s)$ independent
factor in ${\cal A}_{\sigma,\sigma'}^{(v)}$, the amplitude
(\ref{ammain}) is expressed in terms of the correlation functions. They
are calculated from the integral (\ref{ammain}) at vanishing
$(\lambda_s,\bar\lambda_s)$.
In so doing the
linear sources of the
fields and of the global variables are added to the exponent
(it is the known trick in the calculation of correlation functions).
From (\ref{ghact})
and (\ref{lact}), it follows that $c\equiv c(z,\bar z)$ and
$\varsigma_m(z,\bar z)$ are combined into the $\tilde c(z,\bar z)$
field where $\tilde c(z,\bar z)=c(z,\bar z)+ \sum_s\varsigma_m(z,\bar
z)\Lambda_m^{b}$.  The $<\tilde cb>$ correlator (at $\lambda_s=0$) is
\cite{danphr,dan04,dannph}
\begin{equation} <\tilde c(z,\bar z)
b(z',\bar z')>=G_b(z,z')
\label{gbcor}
\end{equation}
where
$G_b(z,z')\to1/(z-z')$ at $z\to z'$, and
\begin{eqnarray}
G_b(g_s(z),z')=\frac{\partial g_s(z)}{\partial z}G_b(z,z')+
\sum_{m=1}^{(3(g-1)}\frac{\partial g_s(z)}{\partial q_m}\chi_m(z')\,,
\label{trangb} \\
\chi_m(z)=-<b(z,\bar z)\Lambda_v^b>
\label{nutm}
\end{eqnarray}
The last term on the right side of the first equation in (\ref{trangb})
appears due to the discontinuity
(\ref{jump}) of $\varsigma_m$.  Furthermore,
$G_b(z,z')$ is not changed under rounds about $A_s$ -cycles.
As the function of $z'$, the
correlator (\ref{gbcor}) is the conform 2-rank tensor.   These
properties are sufficient to determine both $G_b(z,z')$ and
2-rank-tensor zero modes
$\chi_m(z)$.
Henceforth the amplitude (\ref{ammain}) does not depend on
details of $\varsigma_m(z,\bar z)$. It should be noted that
$G_b(z,z')$ is given \cite{dan04,dannph} by
a Poincar\'e series that is not expressed through a local
combination of theta-like functions.

In \cite{ver} the correlator (\ref{gbcor})  is
mistakenly replaced by
$<c(z,\bar
z)b(z',\bar z')>\equiv G_b(z,z';p)$
depending on $3(g-1)$ arbitrary points $p=\{p_m\}$.
The $G_b(z,z';p)$ correlator is one-valued on the Riemann surface,
has poles at $z=p_m$ and  vanishes at
$z'=p_m$. This properties determine $G_b(z,z';p)$ up no a numerical
factor.  In our normalization of the fields the
correlators are
related, as  follows
\begin{equation}
G_b(z,z';p)=G_b(z,z')-\sum_mG_b(z,p_a)\tilde\chi_m(z')\,,\quad
\tilde\chi_m(p_n)=\delta_{mn}
\label{vcor}
\end{equation}
where $\tilde\chi_m(z)$ are the 2-rank tensor zero modes which are
normalized as it shown in (\ref{vcor}).
Due to eq.(\ref{trangb}), the right side of (\ref{vcor})
is one-valued on the Riemann surface. In addition, it
has poles at $z=p_a$,  vanishes at
$z'=p_a$ and goes to $1/(z-z')$ at $z\to z'$.
Thus the right
side of (\ref{vcor}) coincides with $G_b(z,z';p)$.

To  clarify discrepancy between the amplitude (\ref{ammain}) and
the corresponding amplitude in \cite{ver} we
transform (\ref{ammain}) to an integral where the $(b,\bar b)$ fields
each vanishes in $3(g-1)$ points on the Riemann surface.
For this aim the $(\varsigma_m,\overline\varsigma_m)$
functions
are properly specified, and the proportional to $\beta\Lambda_m^b$
and to  $\bar\beta\overline\Lambda_m^b$
terms are removed from the exponent in (\ref{ammain}) by a relevant
shift of the $(\gamma,\bar\gamma)$ fields.  In more details,
$\gamma\to\gamma+ \sum_m\tilde\varsigma_m\Lambda^b_m$ where
$\tilde\varsigma_m\equiv\tilde\varsigma_m(z,\bar z;p)$
and $\varsigma_m\equiv\varsigma_m(z,\bar z;p)$
depend on the $p=\{p_a\}$ set of
$3(g-1)$ points $p_a$ on the Riemann surface.
Furthermore,
\begin{eqnarray}
\tilde\varsigma_m(z,\bar z;p)=-\frac{1}{\pi}\int
G_\sigma(z,z';\{\phi\})[
\partial_{z'}\phi(z',\bar
z')\,\varsigma_m(z,\bar z;p)-\frac{1}{2}
\phi(z',\bar z')\partial_{z'}\varsigma_m(z,\bar z;p)
\nonumber\\
+\partial_{q_m}\phi_-(z,\bar z)]d^2z'\,,
\nonumber\\
\varsigma_m(z,\bar z;p)=\sum_aG_b(z,p_a)\widehat N_{am}
+\frac{1}{2\pi}\int G_b(z,z')\phi(z',\bar z')
\tilde\varsigma_m(z,\bar z;p)d^2z'\,,
\nonumber\\
\widehat N_{am}=\sum_n\widetilde N_{an}^{-1}{\cal N}_{nm},\quad
\widetilde N_{na}=\chi_n(p_a)\,,
\nonumber\\
{\cal N}_{nm}+\frac{1}{2\pi}\int\chi_n(z')\phi(z',\bar z')
\tilde\varsigma_m(z,\bar z;p)d^2z'=1
\label{verl}
\end{eqnarray}
where $\widetilde N_{an}^{-1}$  is the element of the
$\widetilde N^{-1}$ matrix inversed to the
matrix $\widetilde N$.
The 2-rank-tensor
zero modes $\chi_n(z)$  are the same as in (\ref{nutm}). The
$G_\sigma(z,z';\{\phi\})$ function satisfies to the equation as follows
\begin{equation}
\partial_{\bar
z}G_\sigma(z,z';\{\phi\})=\pi\delta^2(z-z')-\sum_m\phi_{-m}(z,\bar z)
\widehat\chi_r^f(z')\,,\quad \int\widehat\chi_r^f(z)\phi_{-n}(z,\bar
z)d^2z=\delta_{mn}
\label{phicor}
\end{equation}
where $\delta_{mn}$ is the Kronecker symbol and
$\widehat\chi_r^f(z')$ is 3/2-rank-tensor zero modes.
In this case
$G_\sigma(z,z';\{\phi\})=-<\gamma(z,\bar z)\, \beta(z',\bar
z')>$ where  $<\gamma\beta>$ is the correlator
at vanishing Grassmann moduli. Due to the
$\sim\Lambda^f_s$ terms in (\ref{lact}), it is a functional of
$\phi_{s-}$. The $(\bar\gamma,\bar\beta)$-dependent
part of the exponent
is represented in the kindred manner.
The $(\gamma,\beta)$ and
$(\gamma,\beta)$ correlators at arbitrary $\phi_{\mp m}$ are expressed
through the correlators in the case when $\phi_{-m}$
and $\phi_{s+}$ each is localized
at  $z=r_s$ and, respectively, at $z=r'_s$, as follows
\begin{equation}
\phi_{s-}(z,\bar z)=\delta^2(z-r_s)\,,\quad
\phi_{s+}(z,\bar z)=\delta^2(z-r'_s)\,.
\label{phidel}
\end{equation}
In this case we denote the $(\gamma,\beta)$ correlator as
$<\gamma(z,\bar z)\, \beta(z',\bar
z')>=-G_\sigma(z,z';r)$
where $r=\{r_s\}$. This correlator
was calculated in \cite{ver}. It
has the pole  at $z=r_j$,
the residue being the 3/2-rank-tensor zero mode $\chi_s^f(z')$
satisfying to the condition
$\chi_s^f(r_j)=-\delta_{sj}$. Then
\begin{equation}
G_\sigma(z,z';\{\phi\})=
G_\sigma(z,z';r)+\sum_s\frac{1}{\pi}\int
G_\sigma(z,z_1;r)\phi_{s-}(z_1,\bar
z_1)d^2z_1\widehat\chi_s^f(z')\,,
\label{corrf}
\end{equation}
\begin{equation}
\chi_s^f(z)+\frac{1}{\pi}
\sum_r\int
\chi_s^f(z')\phi_{r-}(z',\bar z')d^2z'\widehat\chi_r^f(z)=0\,.
\label{hphif}
\end{equation}
Indeed, one can verify that the right side of (\ref{hphif}) satisfies
to (\ref{phicor}). The kindred expression exists for the
$(\bar\gamma,\bar\beta)$ correlators. Once
the integration over the global variables being performed,
the vacuum amplitude ${\cal
A}_{\sigma,\sigma'}^{(v)}$ is found to be as follows
\begin{equation} {\cal
A}_{\sigma,\sigma'}^{(v)}=\int(D{\cal F})\, W_{R\sigma}(p,\bar
p)\,W_{L\sigma'}(p',\bar p') \exp[S_m+ S_{gh}^R+S_{gh}^L]
\label{vamain}
\end{equation}
where
\begin{equation}
W_{R\sigma}(p,\bar p)=\frac{\det {\cal N}}{\det[\chi_m(p_a)]}
\biggl[\prod_{a=1}^{3(g-1)}b(p_a,\bar p_a)\biggl]
\biggl[\prod_{j=1}^{2(g-1)}\delta\biggl(\frac{1}{\pi}\int\beta(z,\bar z)
\phi_{j-}(z,\bar z)\,d^2z\biggl)\biggl]\,,
\label{wfact}
\end{equation}
and $W_{L\sigma'}(p',\bar p')$ is the kindred expression associated
with the left movers. Elements ${\cal N}_{nm}$ of the ${\cal N}$ matrix
is defined in (\ref{verl}). Zero modes $\chi_m(p_a)$ are defined by
(\ref{nutm}).  The product over $a$ in $W_{R\sigma}(p,\bar p)$ provides
the vanishing of the $b(z,\bar z)$ field at $z=p_a$. Eq. (\ref{wfact})
differs from the corresponding expression in \cite{ver} by  the $\det
{\cal N}$ factor (apart from the fact that in \cite{ver}  the set of
2-rank tensor modes is not specified).  This factor generates
the terms in $W_{R\sigma}(p,\bar p)$ which are added to the vacuum
expectations of the supercurrent products (arising from the expanding
of $\exp[S_{gh}^R]$ in $\lambda_j$). As the result, the $G_b(z,z';p)$
correlator is replaced by the $G_b(z,z')$ one. If $\phi_-$ depends on
the Riemann moduli,  additional terms in the amplitude also appear
due to the $\partial_{q_m}\phi_-$ terms in (\ref{lact}).  The
kindred thing arises in the integral over the $(\bar\gamma,\bar\beta)$
fields.

As was noted, the vacuum local amplitude
(\ref{ammain}) contains an
uncertain factor. The factor is independent of the
$(\lambda_s,\bar\lambda_s)$ moduli. Nevertheless,
it depends on the
$\phi_{s\mp}$ fields (\ref{grfield}) because
$S_\Lambda^{R}$ and $S_\Lambda^{L}$ contain
the derivatives of the gravitino field with respect to the
Grassmann moduli.  The factor can be calculated from Ward identities
\cite{dan04}) as was mentioned above. For the commonly used setting
(\ref{phidel}) for $\phi_{s\mp}$ the discussed factor was already
calculated in\cite{ver}. If this setting
is employed, the local vacuum
amplitude is represented as follows
\begin{equation}
{\cal A}_{\sigma,\sigma'}^{(v)}=
\biggl[\frac{1}{\det Im\Omega}\biggl]^5
Z_\sigma(q,r)
\overline{Z_{\sigma'}(q,r')}
Z_{\sigma,\sigma'}^{(mat)}(q,\bar q,
\lambda,\bar\lambda,r,\bar r')
Z_\sigma^{(gh)}(q,
\lambda,r)
\overline{
Z_{\sigma'}^{(gh)}(q,
\lambda,r')}
\label{vcala}
\end{equation}
where $r=\{r_s\}$, $r'=\{r_s'\}$. The
 $(\lambda,\bar\lambda)$-independent  factor
$Z_\sigma(q,r)$ can be taken from
\cite{ver}. The other three factors
differ from the unity only because of  proportional to
$(\lambda,\bar\lambda)$ terms. Among of them,
$Z_{\sigma,\sigma'}^{(mat)}(q,\bar q,
\lambda,\bar\lambda,r,\bar r')$  is due to the expanding in
$(\lambda,\bar\lambda)$ of $\exp S_m$. The last factors are due to
the $(\lambda,\bar\lambda)$ expanding of the rest exponential in
(\ref{ammain}).
The calculation of $Z_\sigma^{(gh)}(q,
\lambda,r)$
is ambiguous because, as it was noted above, the
correlator $<\gamma(z,\bar z)\, \beta(z',\bar z')>=-G_\sigma(z,z';r)$
has  the pole at $z=r_j$.  To resolve the
ambiguity\footnote{In \cite{ver} this matter is treated inexactly.} one
treats fields (\ref{phidel}) as the limit of a spread fields
$\phi_{s\mp}$ using eq.(\ref{corrf}) to calculate of the
$(\gamma,\beta)$ correlator.

\section{Two- and three-loop amplitudes}

In the discussed case the periods
$\Omega_{nm}=\Omega_{mn}$ of the scalar functions
can be taken as moduli. By using eq.
(\ref{sbound}), the integration by parts is performed in
(\ref{main}), and $A_{\sigma,\sigma'}$
is represented as
\begin{equation}
A_{\sigma,\sigma'}=\int {\cal O}(\Omega,\overline\Omega)
{\cal B}_{R\sigma}^T{\cal B}_{L\sigma'}^T
\biggl[{\cal A}_{\sigma,\sigma'}\,
\prod_i\widetilde{\cal
O}(z_i,\bar z_i)\biggl]
d^2\Omega\,d^2\lambda\,
\prod_id^2z_i
\label{rmain}
\end{equation}
where $\widetilde{\cal O}(z_i,\bar z_i)\equiv
\widetilde{\cal O}(z_i,\bar z_i;\Omega,\overline\Omega)$, and
$({\cal B}_{R\sigma}^T,{\cal B}_{L\sigma'}^T)$
are obtained by transposing of
$({\cal B}_{R\sigma},{\cal B}_{L\sigma'})$.  If the
Schottky description is employed, then
\begin{equation}
\widetilde{\cal O}_j(z_j,\bar z_j)=
\rho(1-|g'_s|^2)\rho(1-|\tilde g'_s|^2)
\label{tilb}
\end{equation}
where $g_s'(z)=\partial
g_s(z)/\partial z$,
the $z\to g_s(z)$ transformation is assigned to
$2\pi$-twist about $B_s$-cycle.
The $z\to\tilde g_s(z)$ transformation  is
inverse to the $z\to g_s(z)$ one, so that
$g_s(\tilde g_s(z))= \tilde g_s(g_s(z))=z$.
As above, $\rho(x)$ is the step function.
Instead of the boundary integral in the moduli space,
eq. (\ref{rmain}) contains the integral along the boundary of
the fundamental region of the Klein group.  This integral arises due to
the action of the $({\cal B}_{R\sigma}^T,{\cal B}_{L\sigma'}^T)$
operators on the
$\widetilde{\cal O}(z_i,\bar z_i)$ step function products.  This
boundary integral can be reduced to the integral over the fundamental
region of the Klein group. The above-mentioned reduction is performed
employing the
set of certain functions ${\cal U}_{pq}(z)={\cal U}_{qp}(z)$. The
${\cal U}_{pq}(z)$
function is
unchanged under twists about $A_s$-cycles, but
it has the discontinuity under
$2\pi$-twist $z\to g_s(z)$ about $B_s$-cycle, as follows
\begin{equation}
{\cal U}_{pq}(g_s(z))=\frac{\partial g_s(z)}{\partial
z}{\cal U}_{pq}(z)+ \frac{\partial g_s(z)}{\partial \Omega_{pq}}\,.
\label{tranu}
\end{equation}
The last term on the right side of
(\ref{tranu}) is the  discontinuity of  ${\cal U}_{pq}(z)$.
Then
\begin{equation}
\int{\cal
A}_{\sigma,\sigma'} \frac{\partial\widetilde{\cal O}(z_i,\bar
z_i)}{\partial\Omega_{pq}}\,d^2z_i =\int{\cal A}_{\sigma,\sigma'} {\cal
U}_{pq}(z_i)\frac{ \partial\widetilde{\cal O}(z_i,\bar z_i)}{\partial
z_i}\,d^2z_i\,.
\label{redu}
\end{equation}
Indeed, from
(\ref{tilb}),
the left side and the right side of eq. (\ref{redu})
each is the integral along the $(A_s+\tilde A_s)$ contour where the
$A_s$ contour is given by
the $|g'_s(z)|^2=1$ condition,
and the $\tilde A_s$ contour is given by the $|\tilde g'_s(z)|^2=1$
condition. By the $z\to g(z)$ replacement the integration
along $\tilde A_s$ is reduced to the integration along
$A_s$. The right side of
(\ref{redu}) is calculated using eq. (\ref{trangb}) and taking into
account that ${\cal A}_{\sigma,\sigma'}$ is $(1,1)$-tensor in
$(z_i,\bar z_i)$.  The left side of (\ref{redu}) is calculated using the
relation
\begin{equation}
\frac{\partial\rho(1-|\tilde g_s'(z)|^2)}{\partial
\Omega_{pq}}\biggl|_{z\to g_s(z)}=
-\frac{\partial\rho(1-|g_s'(z)|^2)}{\partial
\Omega_{pq}}+ \frac{\partial\rho(1-|g_s'(z)|^2)}{g_s'(z)\partial z}\,.
\label{difmast}
\end{equation}
As the result, the same expression appears for both the left and
right sides of (\ref{redu})
that proves the validity of eq.(\ref{redu}).  The integration
by parts being performed, eq. (\ref{rmain}) is reduced to  the
integral over the fundamental regions of the modular group and of the
Klein group (there is no any boundary integral), as follows
\begin{equation}
A_{\sigma,\sigma'}=\int \widetilde{\cal A}_{\sigma,\sigma'}
\prod_{m\leq n}d^2\Omega_{mn}\prod_jd^2z_j\,,\quad \widetilde{\cal
A}_{\sigma,\sigma'}=\int {\cal
A}_{\sigma,\sigma'}^{(mod)}\prod_id^2\lambda_i\,,\quad {\cal
A}_{\sigma,\sigma'}^{(mod)}= \widetilde{\cal
B}_{L\sigma'}^T\widetilde{\cal B}_{R\sigma}^T {\cal A}_{\sigma,\sigma'}
\label{rmainf}
\end{equation}
where ${\cal A}_{\sigma,\sigma'}$
is given by (\ref{ammain}).
Operators
$\widetilde{\cal B}_{R\sigma}^T$ and
$\widetilde{\cal B}_{L\sigma'}^T$
are obtained by the
$\partial/\partial\Omega_{pq}\to{\cal D}_{pq}$  replacement
in
${\cal B}_{R\sigma}^T$ and, respectively, by
the $\partial/\partial\overline\Omega_{pq}\to
\overline{\cal D}_{pq}$ replacement in
${\cal
B}_{L\sigma'}^T$. In this case
\begin{equation}
{\cal D}_{pq}=
\frac{\partial}{\partial\Omega_{pq}}-
\sum_i\frac{\partial}{\partial z_i}{\cal U}_{pq}(z_i)\,.
\label{mdop}
\end{equation}
The ${\cal U}_{pq}(z)$  function is constructed
using $G_b(z,z')$  and
zero modes $\chi_n(z)$, see
(\ref{gbcor}),
(\ref{trangb}) and (\ref{nutm}).
In $g=2$ or $g=3$ cases
$n$-index is replaced by a pair $(jl)$ of indices listing
the period matrix elements. Then we use notation $\chi_{(jl)}(z)$
instead of $\chi_n(z)$.
As it is shown below,
\begin{equation}
\chi_{(jr)}(z)=-2\pi iv_j(z)v_r(z)
\label{tzm}
\end{equation}
where
$v_s(z)=\partial_zJ_s(z)$ is 1-form.
For ${\cal U}_{pq}(z)$ one can use
a sum of  $G_b(z,w_j)N^{-1}_{j,(pq)}$ over
$3(g-1)$ arbitrary points $w_j$ where
$N^{-1}$ is the inverse to the $N$ matrix ($\det N\neq0$),
whose matrix elements
$N_{(pq),j}$ are $N_{(pq),j}=\chi_{(pq)}(w_j)$.
This ${\cal U}_{pq}(z)$
depends on $3(g-1)$ arbitrary points $w_j$. For the
calculation  of GSO-projection it
is more convenient to use alternative
functions which
depend on $(2g-3)$ points $w_j$ (these points can be identified with
the 2D-gravitino location points).
These functions are built using 2-rank-tensor
modes
$\zeta_j(u)$,  $\widetilde\zeta_s(u)$ and
$\widehat\tau(u)$ which are
\begin{eqnarray}
\zeta_j(z)=\tau_j(z)-\sum_{s=1}^{g-1}\tau'_j(w_s)
\widetilde\tau_s(z)-[\tau''_j(w_1)-\sum_{s=1}^{g-1}\tau'_j(w_s)
\widetilde\tau''_s(w_1)]\widehat\tau(z)\,,
\nonumber\\
\widetilde\zeta_s(z)=\widetilde\tau_s(z)-\widetilde\tau''_s(w_1)
\widehat\tau(z)
\label{tauf}
\end{eqnarray}
where
2-rank-tensor
modes $\tau_j(z)$ and
$\widetilde\tau_s(z)$ satisfy conditions that
$\tau_j(w_l)=\delta_{jl}$, $\widetilde\tau_s(w_l)$ for all
$w_l$ while $\widetilde\tau_s'(w_l)=\delta_{sl}$ for $l=1$
and (if $g=3$) for $l=2$. In this case, for any $f(x)$ function,
$f'(x)=\partial_xf(x)$ and $f''(x)=\partial_x^2f(x)$.
For $g=2$ one can set  $\tau_1(z)$,
$\widetilde\tau_1(z)$ and $\widehat\tau(z)$, as follows
\begin{equation}
\tau_1(z)=\frac{v_1^2(z)}{v_1^2(w_1)}\,,\quad
\widetilde\tau_1(z)=\frac{v_1(z)d(z,w_1)}{v_1(w_1)d'(w_1,w_1)}\,,
\quad\widehat\tau(z)=
\frac{1}{2}\biggl(\frac{d(z,w_1)}{d'(w_1,w_1)}\biggl)^2
\label{gtwomod}
\end{equation}
where $d'(x,y)=\partial_xd(x,y)$ and
$d(x,y)=v_1(x)v_2(y)-v_1(y)v_2(x)$.
For $g=3$ one can set
\begin{eqnarray}
\tau_1(z)=\frac{d(z,w_2)d(z,w_3)}{d(w_1,w_2)d(w_1,w_3)}\,,\quad
\widetilde\tau_1(z)=\frac{d(z,w_2)d(z,w_1,w_2)}{d(w_1,w_2)
d'(w_1,w_1,w_2)}\,,
\nonumber\\
\widehat\tau(z)=\frac{1}{2}
\biggl(\frac{d(z,w_1,w_2}{d'(w_1,w_1,w_2)}
\biggl)^2
\label{gthrmod}
\end{eqnarray}
where $d(x_1,x_2,x_3)=\det v_j(x_l)$, $d'(x,y,z)=\partial_xd(x,y,z)$
and $d(x,y)=v_1(x)v_2(y)-v_1(y)v_2(x)$. Functions $\tau_2(z)$,
$\tau_3(z)$ and $\widetilde\tau_2(z)$ are obtained by a replacement
of indices in $\tau_1(z)$  and in $\widetilde\tau_1(z)$.
In this case the ${\cal
U}_{nm}(z)$ functions
are defined by
\begin{equation}
-2\pi i{\cal
U}_{nm}(z)=\sum_{j=1}^{2g-3}G_b(z,w_j)\zeta_{j,nm}+
\sum_{s=1}^{g-1}\partial_{w_s}G_b(z,w_s)\widetilde\zeta_{s,nm}
+\partial^2_{w_1}G_b(z,w_1)\widehat\tau_{nm}
\label{unmf}
\end{equation}
where $\zeta_{j,nm}$, $\widetilde\zeta_{s,nm}$ and $\widehat\tau_{nm}$
are coefficients of expansion of
$\zeta_j(u)$,  $\widetilde\zeta_s(u)$ and of
$\widehat\tau(u)$
in the $\chi_{(nr)}(z)/(-2\pi i)$ modes.
Derivatives with
respect to the Riemann moduli
of the correlators and the functions
are
calculated in line with
\cite{danphr,dan04}.
As is proved below,
\begin{equation}
\sum_{m\leq n} \chi_{(mn)}(w)\frac{\partial J_s(z)}{\partial
\Omega_{mn}}= -\frac{\partial R(z,w)}{\partial w }v_s(w)-
v_s(z)G_b(z,w)+\mu_s(w)
\label{moddif}
\end{equation}
where
$R(z,w)$ is the scalar, holomorphic Green function \cite{danphr} for
the $\partial_z\overline\partial_z/\pi$ operator.
This $R(z,w)$ is not
changed  under twists
about  $A_s$-cycles. Under the $2\pi$-twist
$z\to g_s(z)$ about $B_s$ cycle it is changed as follows
\begin{equation}
R(g_s(z),w)=R(z,w)+2\pi iJ_s(w)\,,\quad
R(z,g_s(w))=R(z,w)+2\pi iJ_s(z)\,.
\label{sgf}
\end{equation}
The explicit form of $\mu_s(w)$
is not employed in this paper.
The Green  function $R(z,w)$ differs from
the usual Green function $\ln E(z,w)$ in (\ref{inteq})
only by the scalar zero mode contribution. Thus
\begin{equation}
\partial_z\partial_wR(z,w)=\partial_z\partial_w\ln E(z,w)\,.
\label{sgfpf}
\end{equation}
Differentiating (\ref{moddif}), one obtains that
\begin{equation}
\sum_{m\leq n} \chi_{(mn)}(w)\frac{\partial
v_s(z)}{\partial\Omega_{mn}}= -\frac{\partial^2 R(z,w)}{\partial w
\partial z }v_s(w)- \partial_z[v_s(z)G_b(z,w)]\,.
\label{modonef}
\end{equation}
To derive eq.(\ref{tzm}), the $z\to g_n(z)$ replacement in
(\ref{moddif}) is performed and the relation
$J_s(g_n(z))=J_s(z)+\Omega_{sn}$ is used, along with
eq. (\ref{trangb}) and with the following
relations
\begin{eqnarray}
\frac{\partial J_s(z)}{\partial q_m}\biggl|_{z\to g_s(z)}=
\biggl(\frac{\partial J_s(g_n(z))}{\partial q_m}\biggl)_z-
\frac{\partial g_n(z)}{\partial q_m}v_s(z)\,, \label{difmj}\\
\biggl(\frac{\partial J_s(g_n(z))}{\partial q_m}\biggl)_z=
\frac{\partial J_s(z)}{\partial q_m}+
\frac{\partial \Omega_{sn}}{\partial q_m}\,.
\label{dfmj}
\end{eqnarray}
In eq. (\ref{difmj}) the dervative is calculated under
fixed $z$.

To prove eq. (\ref{moddif}), the difference
$\Delta_s(z,w)$ of the left and right sides of (\ref{moddif}) is
represented as
\begin{equation}
\Delta(z,w)=\oint
\Delta_s(u,w) \partial_uR(u,z)\frac{du}{2\pi i}
\label{oin}
\end{equation}
where the integration is performed along the contour
surrounding the point $u=z$.  The integral (\ref{oin}) is reduced to
the integral along the  boundary (\ref{tilb}) of the integration
region. Further, by the  $z\to g_n(z)$ replacement,
the integral along the
$|\tilde g'_n(u)|^2=1$ contour
is reduced to the integral along the contour
$|g'_n(\tilde g_n(z)|^2=1$.
Then, by using eqs.(\ref{trangb}) and (\ref{dfmj}),
the integral (\ref{oin})  is reduced to the sum of expressions,
each being proportional to the integral of $\partial_uR(u,z)$ along
the $|g'_n(u)|^2=1$ contours.  So far as $R(u,z)$ is unchanged  under
twists about  $A_s$-cycles, the integral vanish that
proves eq. (\ref{moddif}).

The calculation of the same derivatives of
the scalar field correlator
$<x(z_1,\bar z_1)x(z_2,\bar z_2)>=
-X(z_1,\bar z_1;z_2,\bar z_2)$ is simplified, if it is chosen as
follows
\begin{equation}
X(z_1,\bar z_1;z_2,\bar
z_2)=\frac{1}{2}ReR(z_1,z_2)+ \pi \sum_{s,t}ImJ_s(z_1)
\biggl[\frac{1}{Im\Omega}\biggl]_{st}ImJ_t(z_2).
\label{sccor}
\end{equation}
This correlator differs from the scalar field correlator in
\cite{ver} by a scalar zero mode contribution. In the calculation of
the amplitude both the correlators can be used on equal terms.
Further, it can be proved that
\begin{eqnarray}
\sum_{m\leq n}
\chi_{(mn)}(w)\frac{\partial R(z,u)}{\partial \Omega_{mn}}=
-\frac{\partial R(z,w)}{\partial w} \frac{\partial R(w,u)}{\partial
w}\, -\,\frac{\partial R(u,w)}{\partial w}\frac{\partial
R(w,z)}{\partial w}-
\nonumber\\
-G_b(z,w)\frac{\partial R(z,u)}{\partial z}-
-G_b(u,w)\frac{\partial R(z,u)}{\partial u}
+\mu(w)+(z+u)\widehat\mu(w)\,.
\label{grdif}
\end{eqnarray}
An explicit form of $\mu(w)$ and of $\widehat\mu(w)$ is not used
in the paper. The proof of eq. (\ref{grdif}) is similar to the proof of
eq.(\ref{moddif}). From
(\ref{moddif}) and (\ref{grdif}),
it follows that
\begin{eqnarray} \sum_{m\leq n}
\chi_{(mn)}(w)\frac{\partial X(z,\bar z;u,\bar
u)}{\partial\Omega_{mn}}=- 4\widehat X(z,\bar z;w) \widehat X(u,\bar
u;w) -\widehat X(z,\bar z;u)G_b(z,w)
\nonumber\\
-\widehat X(u,\bar
u;z)G_b(z,w)\,,\quad \widehat X(z,\bar z;u)\equiv\partial_u X(z,\bar
z;u,\bar u)\,.
\label{whcor}
\end{eqnarray}
If the relation
\begin{equation}
\sum_{n\leq
m}\chi_{(nm)}(z)\frac{\partial\Upsilon}{\partial\Omega_{nm}} ={\cal
H}(z)
\label{chide}
\end{equation}
takes place for certain
$\Upsilon(z)$, then the derivatives on
its left side are expressed through ${\cal H}(w_j)$ as follows
\begin{equation}
-2\pi i\frac{\partial\Upsilon}{\partial\Omega_{nm}}=
\sum_{j=1}^{2g-3}{\cal H}(w_j)\zeta_{j,nm}+
\sum_{s=1}^{g-1}\partial_{w_s}{\cal H}(w_s)\widetilde\zeta_{s,nm}
+\partial^2_{w_1}{\cal H}(w_1)\widehat\tau_{nm}\,.
\label{dfmf}
\end{equation}
To calculate GSO-projections,
the $r_i$ points (and the $r_i'$ ones) in (\ref{phidel})
are submitted to the conditions
\begin{equation}
\sum_{i=1}^{2g-2}{\bf r}_i-2{\bf\Delta}=0
\label{jac}
\end{equation}
where ${\bf\Delta}$ denotes the vector of the Riemann constants and
${\bf r}_i$ is related to $r_i$ by the Jacobi mapping .
Any
$(g-1)$ points  in the  $\{r_i\}$ set can be taken
at will.
It follows from (\ref{jac}) that
\begin{equation}
d(r_{p_1},\dots,r_{p_g})\equiv\det[v_s(r_{p_j})]=0\,, \quad
v_g(r_{p_i})=\sum_{s=1}^{g-1}\alpha_sv_s(r_{p_i})
\label{detr}
\end{equation}
where $\alpha_s$ are the same for every
$r_{p_i}\subset\{r_i\}$.
Furthermore,
\begin{equation}
d(r_{s_1},\dots,r_{s_{g-1}},z)=f(z;g)d_{gg}(r_{s_1},\dots,r_{s_{g-1}})\,,
\quad
f(z;g)=v_g(z)-\sum_{s=1}^{g-1}\alpha_sv_s(z)
\label{fdetr}
\end{equation}
where
$d_{gg}(r_{s_1},\dots,r_{s_{g-1}})$ is
$(gg)$-minor of the $d(r_{s_1},\dots,r_{s_{g-1}},z)$ determinant.
Under
(\ref{jac}), the $Z_\sigma(q,r)$ factor in (\ref{vcala}) is simplified
essentially. Besides, the $<\gamma(z,\bar z)\beta(z',\bar z')>$
correlator
can be represented
as follows
\begin{equation}
<\gamma(z,\bar z)\beta(z',\bar z')>=-
\frac{f(z';g)}{f(z;g)}S_\sigma(z,z')\,.
\label{gsig}
\end{equation}
The sums over spin structures (GSO-projections)
are calculated in the known manner using Riemann
relations and Fay identities (see eq. (45) in
Ref. \cite{fay}). The  sums (GSO-projections) of the amplitudes
vanish, if the vertex number $n<4$.
If $n=4$, then only that part of $(V)$ in (\ref{ammain})
contributes to the sum, which contains the product of
all the fermion fields.  Using (\ref{moddif}) and (\ref{whcor}), one
can show that the discussed contribution to
${\cal A}_{\sigma,\sigma'}^{(mod)}$ is
factorized in the $\{r_j\}$ and $\{r_j'\}$ so that
${\cal A}_{\sigma,\sigma'}^{(mod)}$ in (\ref{rmainf})
is replaced by an expression
\begin{equation}
{\cal A}_{\sigma,\sigma'}^{(mod)}\longrightarrow
\biggl[\frac{1}{\det[Im\Omega]}\biggl]^5<V_0>
{\cal A}_\sigma^{(g)}(\Omega,\overline\Omega,\lambda,r,z_V,\bar z_V)
\overline{{\cal A}_{\sigma'}^{(g)}(\Omega,\overline\Omega,
\lambda,r',z_V,\bar z_V)}
\label{amfa}
\end{equation}
with
$z_V=\{z_j\}$, and $<V_0>$ to be the vacuum average of $V_0$ where
\begin{equation}
V_0=\exp[\sum_ji\,k_j\cdot x(z_j,\bar z_j)]\,,\quad
\ln<V_0>=
-\sum_{i<j} k_i\cdot k_j <x(z_i,\bar z_i)x(z_j,\bar
z_j)>\,,
\label{vexp}\\
\end{equation}
$k_j$ being 10-momentum of the $j$-th boson. Further,
terms due to the
differentiating of $\det Im\Omega$ in (\ref{vcala}) are canceled with
that part of $Z_{\sigma,\sigma'}^{(mat)}(q,\bar q,
\lambda,\bar\lambda,r,\bar r')$ in (\ref{ammain}) which arises due to
the last term on the right side of (\ref{sccor}).
As the result, ${\cal
A}_\sigma^{(g)}(\Omega,\overline\Omega,
\lambda,r,z_V,\bar z_V)$  is represented as
\begin{equation}
{\cal
A}_\sigma^{(g)}(\Omega,\overline\Omega,
\lambda,r,z_V,\bar z_V)= \frac{1}{<V_0>}
\widetilde{\cal B}_\sigma^T(\Omega,\lambda,r,z_V)\biggl[<V_0>
\widehat{\cal A}_\sigma^{(g)}(\Omega,\lambda,r,z_V)\biggl]
\label{amfar}
\end{equation}
where
$\widehat{\cal A}_\sigma^{(g)}(\Omega,\lambda,r,z_V)$ is  holomorphic
in its arguments. The
superstring amplitude $A_g$ (that is GSO-projection of (\ref{rmainf}))
at $g=2$ and $g=3$  is given by
\begin{eqnarray}
A_g=\int\biggl[\frac{1}{\det Im\Omega}\biggl]^5 <V_0> {\cal
A}_g(\Omega,\overline\Omega,r,z_V,\bar z_V) \overline{{\cal
A}_g(\Omega,\overline\Omega,r',z_V,\bar z_V)} \prod_{m\leq
n}d^2\Omega_{mn}\,\prod_{j}d^2z_j\,,
\label{srmainf}\\
{\cal
A}_g(\Omega,\overline\Omega, r,z_V,\bar z_V)=\sum_\sigma\int
{\cal A}_\sigma^{(g)}(\Omega,\overline\Omega,
\lambda,r,z_V,\bar z_V)\prod_jd\lambda_j\,.
\label{flamps}
\end{eqnarray}
The calculation of (\ref{flamps}) is simplified drastically, if one
sums up over $\sigma$ before the $\Omega_{mn}$ derivatives will be
taken.  Under conditions (\ref{jac}), there are quite simplified the
$Z_\sigma(q,r)$  factor in (\ref{vcala}).
The dependence on boson polarizations in (\ref{flamps}) is
extracted in the form of the  factor $|K|^2$ that is the same for
$g=2$ and $g=3$, and that is the same as in \cite{hok}.

For the two-loop amplitude the result \cite{hok} is reproduced.
In this case
${\cal
A}_g(\Omega,\overline\Omega, r,z_V,\bar z_V)=
{\cal A}_2(\Omega,\overline\Omega,r,z_V,\bar z_V)$
is found to be
\begin{equation}
{\cal A}_2(\Omega,\overline\Omega,r,z_V,\bar z_V)=
\frac{K}{16\pi^2}
Z_2(\Omega,r,z_V)\Biggl[\widetilde{\cal D}\ln<V_0>+
<\partial x(r_1)>_V\cdot <\partial x(r_2)>_V
+\widehat{\cal
A}_2\biggl]
\label{atwo}
\end{equation}
where $\widehat{\cal A}_2$ is
independent of  boson 10-momenta, $x(r,\bar r)=\{x^M(r,\bar r)\}$
and
\begin{equation}
Z_2(\Omega,r,z_V)=2
\frac{v_1(r_1)v_1(r_2)}{f'(r_1;2)f'(r_2;2)
}\prod_i f(z_i;2)\,,\quad
<\partial x(r)>_V=\frac{<V_0\partial_rx(r,\bar r)>}{<V_0>}.
\label{whzrr}
\end{equation}
In this case $v_1(r)$ is 1-form $v_s(r)$ for $s=1$ and
$f'(r;2)=\partial_r f(r;2)$. The $f(r;2)$ function is defined in
(\ref{fdetr}), and $<V_0\partial_rx(r,\bar r)>$ is the vacuum average
of $V_0\partial_rx(r,\bar r)$.  The $\widetilde{\cal D}$ operator in
(\ref{atwo}) is given by \begin{equation} \widetilde{\cal D}=\frac{\pi
i}{2}\sum_{p\leq q}[v_p(r_1)v_q(r_2)+v_p(r_2)v_q(r_1)]
\biggl[\frac{\partial}{\partial\Omega_{pq}}-
\sum_j{\cal U}_{pq}(z_j)
\frac{\partial}{\partial z_j}\biggl]
\label{tmdop}
\end{equation}
where ${\cal U}_{pq}(z)$ is defined by (\ref{unmf}).
The $\widehat{\cal A}_2$ term in (\ref{atwo}) is actually
equal to zero. The proof of this statement will be presented in a future
publication
where the
summation over spin structures
is planned to  give in details.
The
other terms in (\ref{atwo})  is calculated using (\ref{sccor}),
(\ref{whcor}), (\ref{dfmf}) and (\ref{vexp}). In this case $w_1$
in (\ref{gtwomod}) is chosen among $r_j$, say, $w_1=r_1$, and the
same point $w_1=r_1$ is chosen in (\ref{dfmf}). Then
${\cal A}_2(\Omega,\overline\Omega,r,z_V,\bar z_V)\equiv
\widetilde{\cal A}_2(\Omega,z_V)$ in (\ref{atwo})  is
found to be (below $Z_2\equiv Z_2(\Omega,r,z_V)$ in eq.(\ref{whzrr}))
\begin{equation}
\widetilde{\cal A}_2(\Omega,z_V)=\frac{K}{16\pi^2}
Z_2
\sum_{i,j} k_i\cdot k_j\Biggl[\frac{v_1(r_2)}{v_1(r_1)}\widehat
X(z_i,\bar z_i;r_1)
\widehat
X(z_j,\bar z_j;r_1)
-\widehat X(z_i,\bar z_i;r_1)
\widehat
X(z_j,\bar z_j;r_2)\biggl]\,,
\label{ampf}
\end{equation}
see (\ref{whcor}) and (\ref{vexp}) for notations.
Further, using
eqs.  (\ref{modonef}),
to (\ref{sccor}) and (\ref{dfmf}) (for $g=2$), one can derive some
number of identities due to the fact that the left side of (\ref{dfmf})
is independent of $w_1$.  In the explicit form the identities will be
given in the future publication.  Using these identities, one can
represent (\ref{ampf}) as
\begin{equation}
\widetilde{\cal
A}_2(\Omega,z_V)=\frac{KZ_2}{16\pi^2} \sum_{j,l}k_j\cdot k_l
\frac{[f'(r_2;2)]^2v_1(z_j)
v_1(z_l)v_1(r_1)}{16f(z_j;2)f(z_l;2)v_1^3(r_2)}
+\frac{1}{<V_0>}\sum_j
\frac{f'(r_2;2)}{4f(z_j;2)}\frac{\partial <V_0>}{\partial z_j}
\label{xxtwo}
\end{equation}
where
notations are given in  (\ref{vexp}) and in  (\ref{whzrr}). A
detailed deriving of (\ref{xxtwo}) is planned in the future
publication.  The last sum on the right side of (\ref{xxtwo})
originates in (\ref{srmainf}) terms corresponding globally defined
derivatives with respect to $z_j$. Thus it does not contribute to the
superstring amplitude.  The first sum on the right side of
(\ref{xxtwo}) is calculated using (\ref{whzrr}) and the equation
$f'(r_1;2)/v_1^2(r_1)=- f'(r_2;2)/v_1^2(r_2)$ which is obtained by
differentiating $\det[v_j(r_s)]=0$
with respect to $r_1$. In this calculation one employs that $\partial
r_2/\partial r_1=-v_1(r_1)/v_1(r_2)$. The last equation is obtained by
the differentiation of (\ref{jac}) with respect to $r_1$.
Once the last sum on the
right side being omitted and  the 10-momentum
conservation being  taken into account,
$\widetilde{\cal A}_2(\Omega,z_V)$
becomes to be:
\begin{eqnarray}
\widetilde{\cal A}_2(\Omega,z_V)=\frac{K}{64\pi^2}
\sum_{j\neq l\neq m\neq n}k_j\cdot k_lv_1(z_j)v_1(z_l)v_2(z_m)v_2(z_n)
=\frac{K}{64}Y(\{z,\bar z,k\})\,,
\label{ffrres}\\
Y(\{z,\bar z,k\})=
\frac{1}{6}[{\cal Y}(1,2;3,4)+{\cal Y}(1,3;2,4)+{\cal Y}(1,4;3,2)]\,,
\label{fin}\\
{\cal Y}(i,j;l,m)=(k_i-k_j)\cdot(k_l-k_m)d(z_i,z_j)d(z_l,z_m)
\label{caly}
\end{eqnarray}
where $d(z_i,z_j)=v_1(z_i)v_2(z_j)-v_1(z_j)v_2(z_i)$. The
superstring amplitude $A_2$ is obtained by substitution of
$\widetilde{\cal A}_2(\Omega,z_V)\equiv
{\cal A}_2(\Omega,\overline\Omega,r,z_V,\bar z_V)$
to (\ref{srmainf}). It coincides with the amplitude in
\cite{hok}.

In the three-loop amplitude,
${\cal
A}_g(\Omega,\overline\Omega, r,z_V,\bar z_V)=
{\cal A}_3(\Omega,\overline\Omega,r,z_V,\bar z_V)\equiv {\cal A}_3$ in
(\ref{flamps}) is calculated the kindred manner. Potentially,
${\cal A}_3$ might contain terms of fourth-order, two-order and
of zero-order in  10-momentum components $k^M_j$ of the interaction
bosons.  It can be proved that the fourth-order terms
disappear in (\ref{srmainf}) due to the integration over $z_j$.
It seems plausible that zero-order terms are
absent in ${\cal A}_3$, but it needs a further study. The
calculation of quadratic in $k^M_j$ terms in ${\cal A}_3$
has analogy with
the calculation for $g=2$.  Unlike  the $g=2$ case, these terms
in ${\cal A}_3$ depend on $r_j$ though the whole amplitude
(\ref{srmainf}) is independent of $r_j$. Besides, the
discussed terms in ${\cal A}_3$ contain the $G_b$ function (\ref{gbcor})
that, as it was noted above, can not be expressed in term of a local
combination of theta-like functions. \\ \\
The paper is partially supported by Russian State Grant Scientific
School (RSGSS)-65751. 2010.2

\end{document}